\documentclass[journal=jacsat,manuscript=article]{achemso}
\usepackage{amsmath}
\usepackage{algorithm}
\usepackage{algpseudocode}
\usepackage{bm}

\usepackage[version=3]{mhchem} 

\usepackage[hidelinks]{hyperref}
\hypersetup{colorlinks=false}

\usepackage{xcolor}

\usepackage{xr}
\usepackage[nameinlink, capitalise]{cleveref}

\usepackage{threeparttable}

\newcommand{\beq}{\begin{equation}}
\newcommand{\eeq}{\end{equation}}
\newcommand{\beqa}{\begin{eqnarray}}
\newcommand{\eeqa}{\end{eqnarray}}

\newcommand{\binp}[2]{\left( #1 \mid #2 \right)}

\newcommand{\ket}[1]{\mid\! #1 \rangle}

\newcommand{\exptat}[3]{\langle #1 \mid #2 \mid #3 \rangle}
\newcommand{\diraca}[2]{\langle #1 \mid  #2 \rangle}

\newcommand{\propd}[3]{\langle \langle #1 ; #2 \rangle \rangle_{#3}}

\newcommand{\Kap}[1]{\kappa^{ab}_{ij}(#1)}
\newcommand{\Kapg}[2]{\kappa^{#1}_{#2}}

\newcommand{\Del}[1]{\delta_{#1}}

\newcommand{\mptcoef}[2]{\tilde{\kappa}^{#1}_{#2}}

\newcommand{\adag}{\hat a^\dagger}
\newcommand{\aan}{\hat a}

\author{Lina Krarup Knudsen}
\affiliation[UCPH]
{Department of Chemistry, University of Copenhagen, Universitetsparken 5, 2100 Copenhagen, Denmark}
\altaffiliation{Current address: Department of Energy Conversion and Storage, Technical University of Denmark, Agnes Nielsens Vej 301, 2800 Kgs. Lyngby, Denmark}
\email{lkrkn@dtu.dk}

\author{Annika Weisberg Eenholt} 
\affiliation[UCPH]
{Department of Chemistry, University of Copenhagen, Universitetsparken 5, 2100 Copenhagen, Denmark}

\author{Simone Leigh Engell Robinson} 
\affiliation[UCPH]
{Department of Chemistry, University of Copenhagen, Universitetsparken 5, 2100 Copenhagen, Denmark}

\author{Juliane H. Fuglsbjerg} 
\affiliation[UCPH]
{Department of Chemistry, University of Copenhagen, Universitetsparken 5, 2100 Copenhagen, Denmark}

\author{Kasper F. Schaltz} 
\affiliation[UCPH]
{Department of Chemistry, University of Copenhagen, Universitetsparken 5, 2100 Copenhagen, Denmark}
\altaffiliation{Current address: Department of Chemistry, Technical University of Denmark, Kemitorvet 206, 2800 Kgs. Lyngby, Denmark}

\author{Lilli Irene Ør Kristensen} 
\affiliation[UCPH]
{Department of Chemistry, University of Copenhagen, Universitetsparken 5, 2100 Copenhagen, Denmark}

\author{Jens Oddershede} 
\affiliation[SDU]
{Department of Physics, Chemistry and
Pharmacy, University of Southern Denmark, DK-5230
Odense M, Denmark}

\author{Stephan P. A. Sauer}
\affiliation[UCPH]
{Department of Chemistry, University of Copenhagen, Universitetsparken 5, 2100 Copenhagen, Denmark}
\email{sauer@chem.ku.dk}

\title[An \textsf{achemso} demo]
  {SOPPA with the Complete Third-Order A Matrix: Singlet Excitation Energies}

\abbreviations{SOPPA, TOPPA}

\begin{document}


\begin{abstract}

We present the derivation and implementation of the complete third-order correction of the A matrix in the Møller-Plesset polarization propagator approximation. A new contribution to the third-order correction was discovered and we thus report the first results obtained with the complete third-order \textbf{A} matrix. The performance of the Second Order Polarization Propagator Approximation including the third order \textbf{A} matrix (SOPPA-A3) method in calculation of vertical excitation energies was assessed by comparing to the reference methods FCI or CCSDTQP. A substantial improvement over SOPPA was found with the inclusion of the complete third-order \textbf{A} matrix as the SOPPA-A3 method yields excitation energies of a quality comparable to CC2. 
The inclusion of the new contribution results in a change in the individual excitation energies of varying size and direction depending on the molecule and state and the contribution should therefore be included for a formally complete third-order treatment of the \textbf{A} matrix.



\end{abstract}


\section{Introduction}
Accurate predictions of frequency dependent molecular response properties, transition moments and excitation energies are essential within the field of chemistry. 
Computational chemistry provides a range of methods for the calculation of such quantities, each with its own balance of accuracy and computational costs. 
On one end of the spectrum are Coupled Cluster \cite{WCMS-2-566-2012-Sneskov, spas228} and the multi-configurational methods \cite{CPC-13-28-2012-Gonzalez}, which provide high accuracy but come with high computational costs. 
A low-cost method is the Time-Dependent Density Functional Theory (TD-DFT).\cite{Casida2009} 
However, TD-DFT requires a functional optimized for both the property, type of molecule and basis set, and offers no guarantee for better approximations if a more complex functional is employed. 
An alternative is the Second-Order Polarization Propagator Approximation (SOPPA)\cite{Oddershede_1977_particle-hole,ExplicitA2for1980paper, SOPPA1980, SOPPAImplementationInDALTON1996, spas037, spas188}, which builds on a Møller-Plesset (MP) perturbation theory \cite{MPPT, mbpt76pbs} expansion of the ground state wavefunction. 
For small to medium size molecules, this method has been found to yield oscillator strengths, frequency dependent polarizabilities and nuclear spin-spin coupling constants of higher accuracy than CC2 \cite{spas095, spas150, spas190}, and with lower computational costs. 
However, SOPPA tends to underestimate excitation energies by about 0.5 eV \cite{CHRISTIANSEN199847, spas087, spas150}, and is therefore typically not used for calculations of excitation energies.

One way to improve the accuracy of SOPPA is to replace the first-order doubles and second-order singles MP correlation coefficients by Coupled Cluster Singles and Doubles (CCSD) amplitudes leading to the SOPPA(CCSD) method.\cite{jod55, TOPPA1991, spas023}
This, however, requires solving the non-linear CCSD amplitudes.
An alternative way to improve the accuracy of SOPPA excitation energies is to include higher order MP perturbation theory corrections to the matrices in the SOPPA formulation of the response function, \textit{i.e.} to go to the Third-Order Polarization Propagator Approximation (TOPPA). 
A first step in this direction was made by Oddershede and Jørgensen \cite{Oddershede_1977_particle-hole} who presented the equations for the first three contributions to the third-order correction, $\mathbf{A}^{(3)}$, and Eriksen and co-workers \cite{Eriksen, TOPPA1991}, who added the contributions to the $\mathbf{A}$ matrix when computing the seven lowest singlet and triplet excitation energies of beryllium. 
This method showed an improved accuracy compared to SOPPA, with the absolute mean deviation from the full configuration interaction (FCI) results reduced by $0.3$ eV. The standard deviation was found to decrease as well.
However, the equations for the $\mathbf{A}^{(3)}$ matrix, and the calculations themselves, did not include a third-order contribution analogous to the orbital-energy term introduced at second order in the original SOPPA equations,\cite{SOPPA1980} which is not always mentioned in later publications.\cite{ODDERSHEDE1984}

In this work, we present an expression for the general $n$'th order correction of the spin-adapted $\mathbf{A}$ matrix expressed in terms of one- and two-electron reduced density matrices (RDMs). 
This allows us to consistently re-derive the third-order expression
, where we identify a missing contribution and thereby obtain the first complete singlet and triplet third-order $\mathbf{A}$ matrix expressions.



To assess the performance of SOPPA augmented by the complete $\mathbf{A}^{(3)}$ matrix, singlet excitation energies are computed for two sets of in total 10 small molecules and 1 atom. The results are compared to the results using the method introduced by Eriksen and co-workers, SOPPA augmented by the incomplete $\mathbf{A}^{(3)}$ matrix, where the contribution containing the orbital energies is missing. In addition, the excitation energies are compared to SOPPA, CC2, CCSD excitation energies and as reference either FCI or Coupled Cluster singles, doubles, triples, quadruples and pentuples (CCSDTQP) results.

The paper is organized as follows. First we review the theory of MP perturbation theory based polarization propagators. Then we present the $n$'th order $\mathbf{A}$ matrix expression and derive the equations for the second- and third-order $\mathbf{A}$ matrix. Finally, we present the results for the test sets of molecules and draw conclusions about the performance of the SOPPA-A3 method.

\section{Theory}
\subsection{The Polarization Propagator}
Ground-state linear response properties, transition moments, and excitation energies can be obtained from the linear response function also called the polarization propagator.\cite{YellowBook,spasB8} 
In its matrix representation, the polarization propagator of the number-conserving, one-electron operators $\hat{P}$ and $\hat{O}$ in the frequency domain is given by
\beq
\propd{\hat{P}}{\hat{O}}{\omega} = \mathbf{T}^{T}(\hat{P}) ~ (\hbar \omega \mathbf{S}-\mathbf{E})^{-1} ~\mathbf{T}(\hat{O})
\label{PP}
\eeq
where $\mathbf{T}^{T}(\hat{P})$ and $\mathbf{T}(\hat{O})$ are the property gradient vectors, $\textbf{S}$ is the metric matrix, and $\textbf{E}$ is the electronic Hessian. 
The aforementioned matrices are defined as
\beqa
    S_{ij} &=& \exptat{0 }{ [ \hat{h}^{\dagger}_{i} , \hat{h}_{j} ] }{0}
    \label{S_matrix}
\\
    E_{ij} &=& \exptat{0}{ [ \hat{h}^{\dagger}_{i} ,[ \hat{H}, \hat{h}_{j} ]] }{ 0}
    \label{E_matrix}
\eeqa
where $|0\rangle$ is the unperturbed ground state, $\hat{H}$ is the field-free Hamiltonian in the Born– Oppenheimer approximation, and $\{ \hat{h}_{n} \}$ is a complete set of excitation and de-excitation operators.
To determine the excitation energies, the poles of the polarization propagator are determined from the condition
\beq 
(\mathbf{E}-\hbar \omega \mathbf{S}) \mathbf{X} = 0, \label{EigenvalueProblem-1} 
\eeq
which is the generalized eigenvalue problem
\beq 
\mathbf{E} \mathbf{X} = \hbar \omega \mathbf{S} \mathbf{X} ~,
\label{EigenvalueProblem} 
\eeq
the solutions of which are the corresponding excitation energies  $\hbar \omega$, and solution vectors,  $\mathbf{X}$.

\subsection{Definition and Structure of SOPPA and TOPPA}
In the MP perturbation theory polarization propagator approximations, the unperturbed ground-state $|0\rangle$, here assumed to be a closed-shell singlet state, is approximated by a MP wavefunction. The field-free electronic Hamiltonian is partitioned according to MP perturbation theory \cite{MPPT}
\beq
\hat{H}=\hat{F}+\hat{V},
\label{eq6}
\eeq
where $\hat{F}$ is the Fock operator and $\hat{V}$ the fluctuation potential. In a basis of canonical spatial orbitals $\{\phi_p(\boldsymbol{r})\}$ the Fock operator\cite{PinkBible} is given as
\beqa
\hat{F} &=& \sum_{p} \epsilon_{p} \hat{E}^{S}_{pp} ,
\eeqa
where $\epsilon_p$ is the orbital energy of $\phi_p$ and the singlet spin-adapted excitation operators are
\beq
\hat{E}^{S}_{pq} =  \adag_{p\alpha}\aan_{q\alpha} + \adag_{p\beta}\aan_{q\beta} .
\eeq
In the expression, $\adag_{p\alpha}$ and $\aan_{q\alpha}$ are creation and annihilation operators, respectively.\cite{PinkBible} 
Inserting the expression for the Fock operator of a closed-shell system and the non-relativistic and spin-free electronic Hamiltonian in \cref{eq6}, 
the fluctuation potential becomes\cite{PinkBible}
\beqa
\hat{V} &=& \hat{H} - \hat{F} \\ &=& \frac{1}{2}\sum_{pqrs} \binp{pq}{rs} { \hat{E}^{S}_{pq}  \hat{E}^{S}_{rs} - \frac{1}{2}\sum_{pqs}  \binp{pq}{qs} \hat{E}^{S}_{ps} }
- \sum_{pq i} \left \{ 2 \binp{pq}{ii} - \binp{pi}{iq} \right \} \hat{E}^{S}_{pq} ,
\eeqa
where $\binp{pq}{rs}$ is the two-electron integral (in atomic units) over spatial orbitals in the Mulliken notation
\begin{equation}
\binp{pq}{rs}= \iint \frac{\phi^{*}_p(\boldsymbol{r}_1)\phi^{*}_r(\boldsymbol{r}_2)\phi_q(\boldsymbol{r}_1)\phi_s(\boldsymbol{r}_2)}{r_{12}}d\boldsymbol{r}_1d\boldsymbol{r}_2 .
\end{equation}
Here, and in the following, $p, q, r, s, \cdots$ denote general orbital indices, $i, j, k, l, m, \cdots$ refer to orbitals occupied in the Hartree-Fock reference $\ket {\textrm{HF}}$, and $a, b, c, d, e, \cdots$ to unoccupied orbitals.

The MP wavefunction consists of corrections to the Hartree-Fock wavefunction of increasing order in the fluctuation potential
\begin{equation}
\ket {\textrm{MP}} = \ket {\textrm{HF}} + \ket{0^{(1)}} + \ket{0^{(2)}} + \cdots \,.
\end{equation}
Only doubly excited determinants contribute to the first order correction
\beq
 \ket{0^{(1)}} = \ket{0_{\textrm{D}}^{(1)}} =  \frac{1}{2} \sum _{i j} \sum _{a b} ~
 \hat{E}^{S}_{ai} \hat{E}^{S}_{bj} \ket {\textrm{HF}} ~ ^{(1)}\Kapg{ab}{ij} 
 \label{DE1}
 \eeq
whereas singly to quadruply excited determinants give contributions to the second order correction
\beq
    \ket{0^{(2)}} = \ket{0_{\textrm{S}}^{(2)}} + \ket{0_{\textrm{D}}^{(2)}} + \ket{0_{\textrm{T}}^{(2)}} + \ket{0_{\textrm{Q}}^{(2)}} .
\eeq
However, for both SOPPA and TOPPA, the complete set of operators $\{\hat{h}_{n}\}$ is approximated by single and double excitation and de-excitation operators relative to the Hartree-Fock reference, and therefore it is only necessary to include up to doubly excited determinants in the wavefunction
%
%
\beqa
\ket{0_{\textrm{S}}^{(2)}} &=& \frac{1}{\sqrt{2}} ~ \sum _{i a} ~  \hat{E}^{S}_{ai} \ket {\textrm{HF}}  ~ ^{(2)}\Kapg {a}{i} ,
 \label{SE2}
\\
 \ket{0_{\textrm{D}}^{(2)}} 
 &=& \frac{1}{2} \sum _{i j} \sum _{a b} ~
 \hat{E}^{S}_{ai} \hat{E}^{S}_{bj} \ket {\textrm{HF}} ~ ^{(2)}\Kapg{ab}{ij}. 
 \label{DE2}
 \eeqa
%
%
The first and second order doubles correlation coefficients are defined in \cref{eq:first_order_kappa_2,eq:second_order_kappa_2}, respectively, \cite{MPPT,Oddershede_1977_particle-hole,ODDERSHEDE1984,TOPPA1991,spas216}
%
%
%
%
\begin{align}
    ^{(1)}\kappa^{ab}_{ij} = - \Delta^{ab}_{ij}  (ia|jb) ,
    \label{eq:first_order_kappa_2}
\end{align}
\begin{align}
    ^{(2)}\kappa^{ab}_{ij} =& - \hat{P}_{ij}^{ab} \Delta^{ab}_{ij} \bigg( \frac{1}{2} \sum_{cd} (ac|bd) ^{(1)}\kappa^{cd}_{ij} + \frac{1}{2} \sum_{kl} (ij|kl) ^{(1)}\kappa^{ab}_{kl} \nonumber\\
    & \qquad \qquad \quad + 2 \sum_{kc} (ia|kc) ^{(1)}\kappa^{cb}_{kj} - \sum_{kc} (ik|ac) ^{(1)}\kappa^{cb}_{kj} \nonumber\\
    & \qquad \qquad \quad - \sum_{kc} (ik|bc) ^{(1)}\kappa^{ac}_{kj} - \sum_{kc} (jb|kc) ^{(1)}\kappa^{ca}_{ik} \bigg).
    \label{eq:second_order_kappa_2}
\end{align}
Here $\Delta^{ab}_{ij}$ is the orbital energy difference denominator
\begin{align}
    \Delta^{ab}_{ij} = \frac{1}{\varepsilon_a - \varepsilon_i + \varepsilon_b - \varepsilon_j},
    \label{eq:orb_energy_denom}
\end{align}
and $\hat{P}_{ij}^{ab}$ is a permutation operator that acts on a four index quantity as
\begin{align}
    \hat{P}_{qs}^{pr} A_{qs}^{pr} = A_{qs}^{pr} + A_{sq}^{rp} ,
\end{align}
to preserve permutational symmetry.
The second order singles correlation coefficients can be expressed using modified first order doubles correlation coefficients
\beq
^{(2)}\Kapg{a}{i} = \frac{1}{\sqrt{2}} ~
\frac{1}{\epsilon_i - \epsilon_{a}}
\left ( \sum_{jbc} \binp{ab}{jc} ~ ^{(1)}\mptcoef{bc}{ij}
- \sum_{jkb} \binp{ki}{jb} ~ ^{(1)}\mptcoef{ba}{jk} \right ) ,
\eeq
which are defined as a ``Coulomb minus Exchange'' like combination of the following form
\beqa
^{(1)}\mptcoef{ab}{ij} &=& 4 ^{(1)}\Kapg{ab}{ij} - 2 ^{(1)}\Kapg{ab}{ji}.
\eeqa
%
%
%
%
%
%
%
The polarization propagator in the matrix representation, Eq. (\ref{PP}), can be written in a partitioned form\cite{spasB8}, which is then evaluated to second- or third-order in the fluctuation potential for SOPPA or TOPPA, respectively.
%
%
The resulting block structured Hessian matrices for SOPPA and TOPPA are 
\beqa
\mathbf{E}_{\textrm{SOPPA}}&=&
\begin{pmatrix}
\mathbf{A}^{(0,1,2)} & {\mathbf{B}^{(1,2)*}} & {\tilde{\mathbf{C}}^{(1)}} & \mathbf{0}\\
\mathbf{B}^{(1,2)} & {\mathbf{A}^{(0,1,2)*}} & \mathbf{0} & {\tilde{\mathbf{C}}^{(1)*}} \\
\mathbf{C}^{(1)} & \mathbf{0} & \mathbf{D}^{(0)} & \mathbf{0} \\
\mathbf{0} & {\mathbf{C}^{(1)*}} & \mathbf{0} & {\mathbf{D}^{(0)*}}
\end{pmatrix} ,
\eeqa
and in TOPPA
\beqa
\mathbf{E}_{\textrm{TOPPA}}&=&
\begin{pmatrix}
\mathbf{A}^{(0,1,2,3)} & {\mathbf{B}^{(1,2,3)*}} & {\tilde{\mathbf{C}}^{(1)}} & \mathbf{0}\\
\mathbf{B}^{(1,2,3)} & {\mathbf{A}^{(0,1,2,3)*}} & \mathbf{0} & {\tilde{\mathbf{C}}^{(1)*}} \\
\mathbf{C}^{(1)} & \mathbf{0} & \mathbf{D}^{(0,1)} & \mathbf{0} \\
\mathbf{0} & {\mathbf{C}^{(1)*}} & \mathbf{0} & {\mathbf{D}^{(0,1)*}}
\end{pmatrix} ,
\eeqa
where the superscripts indicate the perturbative order through which the matrix is evaluated, and $\tilde{\bf{C}}^{(1)}$ is the transpose of ${\bf{C}}^{(1)}$, when spin-adapted operators are used, see the section on the $n$'th order $\mathbf{A}$ matrix. The $\mathbf{A}$ and $\mathbf{B}$ matrices are defined as elements of double commutators containing only single excitation and de-excitation operators, Eq. 3, the $\mathbf{D}$ matrix contains only double excitation operators and the $\mathbf{C}$ matrix couples single and double excitations. 

Explicit expressions for matrix elements in the SOPPA method can be found in appendix C in the 1984 paper by Oddershede \textit{et al.} \cite{ODDERSHEDE1984}. Please note that the $\mathbf{A}^{(2)}$ matrix expression is missing the orbital energy term and that Eq. (C.16) has an incorrect sign\cite{SOPPAImplementationInDALTON1996}.

A complete TOPPA method will thus require  updating the SOPPA matrices with third order contributions to the A and B matrices as well as an additional D(1) contribution. Based on the relative importance of this matrices in the SOPPA method it is our expectation  that A(3)  will give the most important third order term and this is therefore the contribution we will focus on the rest of this publication.
In order to re-derive expressions for the $\mathbf{A}^{(3)}$ matrix, we will in the following section first present a generic expression for the $n$'th order $\mathbf{A}$ matrix.

\section{The $n$'th order A matrix}

The zeroth, first and second order expressions for element $ai,bj$ of the $\mathbf{A}$ matrix of the MP polarization propagator approximation \cite{Oddershede_1977_particle-hole,ODDERSHEDE1984} can be generalized to a (normalized) $n$'th order expression
\beq
A^{X(n)}_{ai,bj} 
= \left ( \frac{
\exptat{\textrm{MP}}{[{\hat q}^X_{ai},[ \hat F +  \hat V, {\hat{q}_{bj}}^{X\dagger}]]}{\textrm{MP}}} 
{\diraca{\textrm{MP}}{\textrm{MP}}}
\right )^{(n)} ,
\eeq
%
%
%
where the MP wavefunction is included up to the desired order $n$ and $X = S$ or $T$ denotes either singlet or triplet spin-adaptation.
$\hat q_{ai}^{X\dagger}$ and $\hat q_{ai}^X$ are the single excitation and de-excitation operators and in the spin-adapted formulation the excitation operators are defined as
\beqa q_{ai}^{S\dagger} = \frac{1} {\sqrt{2}} \hat{E}^{S}_{ai} 
\eeqa
\beqa
q_{ai}^{T\dagger} = \frac{1} {\sqrt{2}} \hat{E}^{T}_{ai} = \frac{1} {\sqrt{2}} \left ( \adag_{a\alpha}\aan_{i\alpha} - \adag_{a\beta}\aan_{i\beta} \right )
\eeqa
for the singlet and  triplet excitations, respectively, for a closed-shell singlet reference state \cite{ODDERSHEDE1984}. 
In the expression for the \textbf{A} matrix element, the expectation value is divided by the norm of the MP wavefunction, since it is not normalized at a given order $n$.

In contrast to previous derivations of the contributions to the SOPPA or TOPPA matrices,\cite{Oddershede_1977_particle-hole, ExplicitA2for1980paper, SOPPA1980, ODDERSHEDE1984, TOPPA1991, SOPPAImplementationInDALTON1996, spas037, spas188} in this work the commutators are evaluated prior to the perturbative analysis applying the standard second quantization commutator relations. 
This allows us to express the matrix element of the general order $n$ in terms of RDMs.
Keeping only terms of order $n$, one obtains the general element of the singlet or triplet spin-adapted $\mathbf{A}$ matrix, given in terms of normalized one- and two particle RDMs, $\boldsymbol{\rho}$ and $\boldsymbol{\pi}$, which are defined in the following section
\beqa
A^{X(n)}_{ai,bj} &=&
\frac {1}{2} (\epsilon_b - \epsilon_j)
             \left ( \Del{ab} ~\rho^{(n)}_{ij}
                   - \Del{ij} ~\rho^{(n)}_{ba} \right ) 
\nonumber \\
&\phantom{=}&- \left ( \sum _k \binp{ab}{kk}
- \frac{1}{4} \sum _k \binp{ak}{kb}
+ \frac{1}{4} \sum _c \binp{ac}{cb} \right )
\rho_{ij}^{(n-1)}\nonumber \\
&\phantom{=}&- \left ( \sum _k \binp{ji}{kk}
- \frac{1}{4} \sum _k \binp{jk}{ki}
+ \frac{1}{4} \sum _c \binp{jc}{ci} \right )
\rho_{ba}^{(n-1)}\nonumber \\
&\phantom{=}&+ ~\Del{ij} \sum _p \left ( \sum _k \binp{pb}{kk}
- \frac{1}{4} \sum _k \binp{pk}{kb}
+ \frac{1}{4} \sum _c \binp{pc}{cb} \right )
\rho_{pa}^{(n-1)}\nonumber \\
&\phantom{=}&+ ~\Del{ab} \sum _p \left ( \sum _k \binp{jp}{kk}
- \frac{1}{4} \sum _k \binp{jk}{kp}
+ \frac{1}{4} \sum _c \binp{jc}{cp} \right )
\rho_{ip}^{(n-1)}\nonumber \\
&\phantom{=}& + \frac{1}{4} \sum _{pq} \bigg \{
\binp{jp}{qi} \left(  \pi_{bpqa}^{X(n-1)} + \pi_{qabp}^{X(n-1)} \right) +
\binp{ji}{pq} \left( \pi_{bapq}^{S(n-1)} + \pi_{pqba}^{S(n-1)} \right)
\nonumber \\
&\phantom{=}& \phantom{+ \frac{1}{4} \sum _{pq}} +
\binp{ap}{qb} \left( \pi_{ipqj}^{X(n-1)} +  \pi_{qjip}^{X(n-1)} \right)
+ \binp{ab}{pq} \left( \pi_{ijpq}^{S(n-1)} + \pi_{pqij}^{S(n-1)} \right)
\nonumber \\
&\phantom{=}& \phantom{+ \frac{1}{4} \sum _{pq}} - 
\binp{ap}{jq} \left( \pi_{ipbq}^{X(n-1)} + \pi_{bqip}^{X(n-1)} \right)
- \binp{pi}{qb} \left( \pi_{paqj}^{X(n-1)} +  \pi_{qjpa}^{X(n-1)} \right)
\bigg  \} \nonumber \\
&\phantom{=}& - \frac{1}{4} \Del{ij} \sum _{pqr}
\binp{pb}{qr} \left( \pi_{paqr}^{S(n-1)} +\pi_{qrpa}^{S(n-1)}
\right )
- \frac{1}{4} \Del{ab} \sum _{pqr}
\binp{jp}{qr}  \left ( \pi_{ipqr}^{S(n-1)} + \pi_{qrip}^{S(n-1)}
\right ). \nonumber \\
\label{A.n'th.order}
\eeqa

\subsection{Reduced Density Matrices}
The elements of the normalized $n$'th order one-particle RDM, ${\boldsymbol \rho}$, are defined as
\beqa
\rho_{pq}^{(n)} = 
\left ( \frac{
\exptat{\textrm{MP}}{\hat{E}^{S}_{pq}}{\textrm{MP}}} 
{\diraca{\textrm{MP}}{\textrm{MP}}}
\right )^{(n)} ,
\eeqa
and the elements of the singlet and triplet spin-adapted second-order two-particle RDMs, ${\boldsymbol\pi^{X}}$, are given as
\beqa
\pi_{pqrs}^{X(n)} &=& 
\left ( \frac{
\exptat{\textrm{MP}}{\hat{E}^{X}_{pq}  \hat{E}^{X}_{rs}}{\textrm{MP}}} 
{\diraca{\textrm{MP}}{\textrm{MP}}}
\right )^{(n)} .
\eeqa
To first order in the fluctuation potential, there is no correction from the wavefunction normalization, since the Hartree-Fock state is normalized and has no overlap with the other contributions to the MP wavefunction. For second- and third-orders, the denominator of the normalized RDM is expanded as a power series. To obtain an expansion of the RDM to third order the expansion may be truncated after the first order
\beq
\frac{1}{1+\diraca{0^{(1)} + 0^{(2)} + \cdots}{0^{(1)} + 0^{(2)} + \cdots}} = 1 - \diraca{0^{(1)} + 0^{(2)} + \cdots}{0^{(1)} + 0^{(2)} + \cdots}.
\eeq

The second-order contribution to the normalized one-particle RDM has been shown to be\cite{JCP-88-3824-1988-Jensen, JCP-89-5354-1988-Jensen} 
\beqa
    \rho^{(2)}_{ij} & = &-  \sum_{kcd} {}^{(1)}\kappa^{cd}_{ik}  {}^{(1)}\tilde{\kappa}^{cd}_{jk} , \label{D2ij}\\
    \rho^{(2)}_{ab} & = & \sum_{klc}  {}^{(1)}\kappa^{ac}_{kl}  {}^{(1)}\tilde{\kappa}^{bc}_{kl} ,  \label{D2ab}
\eeqa
\beq
\rho^{(2)}_{ia} = \rho^{(2)}_{ai} = {\sqrt{2}} ~^{(2)}\Kapg{a}{i} . \label{D2ai}
\eeq
The third-order contribution to the occupied-occupied (OO) and virtual-virtual (VV) block of the normalized one-particle RDM can be derived to yield
\beqa
    \rho^{(3)}_{ij} & = & \exptat{0_\textrm{D}^{(1)}}{\hat{E}^S_{ij}}{0_\textrm{D}^{(2)}} + \exptat{0_\textrm{D}^{(2)}}{\hat{E}^S_{ij}}{0_\textrm{D}^{(1)}} - \Del{ij} (\diraca{0^{(1)}_\textrm{D}}{0^{(2)}_\textrm{D}} + \diraca{0^{(2)}_\textrm{D}}{0^{(1)}_\textrm{D}}) \nonumber \\
    & = & -  \frac{1}{2} \sum_{kcd} \left( {}^{(1)}\kappa^{cd}_{ik}  {}^{(2)}\tilde{\kappa}^{cd}_{jk} + {}^{(2)}\kappa^{cd}_{ik}  {}^{(1)}\tilde{\kappa}^{cd}_{jk} \right) , \\
    \rho^{(3)}_{ab} & = & \exptat{0_\textrm{D}^{(1)}}{\hat{E}^S_{ab}}{0_\textrm{D}^{(2)}} + \exptat{0_\textrm{D}^{(2)}}{\hat{E}^S_{ab}}{0_\textrm{D}^{(1)}} \nonumber \\
    & = & \frac{1}{2} \sum_{klc} \left( {}^{(1)}\kappa^{ac}_{kl}  {}^{(2)}\tilde{\kappa}^{bc}_{kl} + {}^{(2)}\kappa^{ac}_{kl}  {}^{(1)}\tilde{\kappa}^{bc}_{kl} \right) .
\eeqa
To first order the normalized singlet or triplet spin-adapted two-particle RDM becomes
\beq
\pi^{X(1)}_{pqrs} = \exptat{\textrm{HF}}{\hat{E}^{X}_{pq}\hat{E}^{X}_{rs}}{0_\textrm{D}^{(1)}} + \exptat{0_\textrm{D}^{(1)}}{\hat{E}^{X}_{pq}\hat{E}^{X}_{rs}}{\textrm{HF}}.
\eeq
The only non-vanishing elements of the first-order two-particle RDMs are then the occupied-virtual-occupied-virtual (OVOV) and virtual-occupied-virtual-occupied (VOVO) blocks
\beqa
\pi_{iajb}^{S(1)} &= \pi_{aibj}^{S(1)} &= ~^{(1)}\mptcoef{ab}{ij} , \\
\pi_{iajb}^{T(1)} &= \pi_{aibj}^{T(1)} &=  - 2 ~^{(1)}\kappa_{ji}^{ab}.
\eeqa
The second order correction to the normalized two-particle RDMs are defined as
\beqa 
\pi_{pqrs}^{X(2)} &=& \exptat{0_{\textrm{D}}^{(2)}}{\hat{E}^{X}_{pq}\hat{E}^{X}_{rs}}{\textrm{HF}} 
+ \exptat{\textrm{HF}}{\hat{E}^{X}_{pq}\hat{E}^{X}_{rs}}{0_{\textrm{D}}^{(2)}} \nonumber \\
 &~& + ~ \exptat{0_{\textrm{S}}^{(2)}}{\hat{E}^{X}_{pq}\hat{E}^{X}_{rs}}{\textrm{HF}} 
 + \exptat{\textrm{HF}}{\hat{E}^{X}_{pq}\hat{E}^{X}_{rs}}{0_{\textrm{S}}^{(2)}} \nonumber \\
 &~& + ~ \exptat{0_{\textrm{D}}^{(1)}}{\hat{E}^{X}_{pq}\hat{E}^{X}_{rs}}{0_{\textrm{D}}^{(1)}} 
 - \exptat{\textrm{HF}}{\hat{E}^{X}_{pq}\hat{E}^{X}_{rs}}{\textrm{HF}}  
 \diraca{0_{\textrm{D}}^{(1)}}{0_{\textrm{D}}^{(1)}}.
\eeqa
The elements of this matrix can be split into five different categories based on the number of virtual and occupied orbitals.

\textit{Four virtual indices:} 
In this category there is only one possible matrix element, for which one obtains
\beqa
\pi_{abcd}^{S(2)} &=& \frac{1}{2} \sum_{kl} \big( ~  ^{(1)}\kappa^{ca}_{kl} ~ ^{(1)}\mptcoef{db}{kl} + ~ ^{(1)}\kappa^{ac}_{kl} ~ ^{(1)}\mptcoef{bd}{kl} \big) + \Del{cb} \rho^{(2)}_{ad} , \\
\pi_{abcd}^{T(2)} &=& \delta_{cb} \rho_{da}^{(2)} 
-2 \sum_{kl} ~ ^{(1)}\kappa_{lk}^{ac} ~ ^{(1)}\kappa_{lk}^{db}.  
\eeqa
\textit{Three virtual and one occupied index:} 
In this category there are four possible matrix elements 
\beqa
\pi_{abci}^{S(2)} &=& \pi_{abci}^{T(2)} =  \Del{cb} \rho^{(2)}_{ia} , \\ 
\pi_{iabc}^{S(2)} &=& \pi_{iabc}^{T(2)} =  \Del{ab} \rho^{(2)}_{ic} , \\
\pi_{abic}^{S(2)} &=& \pi_{abic}^{T(2)} = \pi_{aibc}^{S(2)} = \pi_{aibc}^{T(2)} = 0.
\eeqa
\textit{Two virtual and two occupied indices:} 
In this category there are four different matrix elements
\beqa
\pi_{iajb}^{S(2)} &=& \pi_{aibj}^{S(2)} = ~ ^{(2)} \mptcoef{ab}{ij} , \label{eq.pi_iajb}\\
\pi_{iajb}^{T(2)} &=& \pi_{aibj}^{T(2)} =  - 2 ~^{(2)}\kappa_{ji}^{ab} , \label{eq.pi_iajb-t}\\
\pi_{aijb}^{S(2)} &=& \frac{1}{2} ~\sum_{dk} ~  ^{(1)}\mptcoef{ad}{ik} ~ ^{(1)}\mptcoef{db}{kj} , \\
\pi_{aijb}^{T(2)} &=&  2 \sum_{dk } ~ ^{(1)}\kappa_{ik}^{da}~ ^{(1)}\kappa_{jk}^{db} ,
\\
\pi_{abij}^{S(2)} &=& \pi_{ijab}^{S(2)} = - \sum_{dk} \big( ~ ^{(1)}\kappa^{da}_{kj} ~  ^{(1)}\mptcoef{db}{ki} + ~ ^{(1)}\kappa^{ad}_{kj} ~ ^{(1)}\mptcoef{bd}{ki} \big) + 2 ~ \Del{ij} \rho^{(2)}_{ab} , \\
\pi_{abij}^{T(2)} &=& \pi_{ijab}^{T(2)}  = -\sum_{d k} \big( 
 ~ ^{(1)}\kappa_{kj}^{da} ~ ^{(1)} \tilde \kappa_{ki}^{db} 
- 2 ~ ^{(1)}\kappa_{kj}^{ad}~ ^{(1)}\kappa_{ki}^{db} 
\big) , \\
\pi_{iabj}^{S(2)} &=&  
\frac{1}{2} \sum_{ck} \big( ~ ^{(1)}\mptcoef{cb}{kj} ~ ^{(1)}\mptcoef{ac}{ik} \big) +  \Del{ab} \rho^{(2)}_{ij} -  \Del{ij} \rho^{(2)}_{ab} , \\
\pi_{iabj}^{T(2)} &=&    \delta_{ab} \rho^{(2)}_{ji}  - \delta_{ij} \rho_{ba}^{(2)} + 2  \sum_{c k}  ~ ^{(1)}\kappa_{jk}^{cb} ~ ^{(1)}\kappa_{ki}^{ac}. 
\eeqa

\textit{One virtual and three occupied indices:} 
In this category there are four possible matrix elements 
\beqa
\pi_{aijk}^{S(2)} &=& 2 ~\Del{jk} ~ \rho^{(2)}_{ia} ,\\
\pi_{ijka}^{S(2)} &=& 2 ~\Del{ji} ~ \rho^{(2)}_{ka} ,\\
\pi_{aijk}^{T(2)} &=& \pi_{ijka}^{T(2)}  = ~ 0 , \\
\pi_{iajk}^{S(2)} &=& 2 ~ \Del{jk} ~  \rho^{(2)}_{ia} ~ - ~ \Del{ik} ~ \rho^{(2)}_{aj} , \\
\pi_{ijak}^{S(2)} &=& 2 ~ \Del{ji} ~  \rho^{(2)}_{ka} ~ - ~ \Del{ik} ~ \rho^{(2)}_{aj} , \\
 \pi_{iajk}^{T(2)} &=& \pi_{ijak}^{T(2)}  = ~ - ~ \delta_{ik} ~ \rho^{(2)}_{aj}.
\eeqa

\textit{Four occupied indices:}
In the final category there is only one possible matrix element, which is given as
\beqa
\pi_{ijkl}^{S(2)} &=&  
- \Del{li} \rho^{(2)}_{jk} +  2 \Del{ij} \rho^{(2)}_{lk} + 2 \Del{kl} \rho^{(2)}_{ji}  + \frac{1}{2} \sum_{cd} \big( ~ ^{(1)}\kappa^{cd}_{lj} ~ ^{(1)}\mptcoef{cd}{ki} + ~ ^{(1)}\kappa^{cd}_{jl} ~ ^{(1)}\mptcoef{cd}{ik} \big) , \\
\pi_{ijkl}^{T(2)} &=& - \delta_{li} \rho_{jk}^{(2)} - 2 \sum_{cd} ~ ^{(1)}\kappa_{jl}^{cd} ~ ^{(1)}\kappa_{ki}^{cd}.
\eeqa

\subsubsection{The Second-Order Correction to the A matrix}
The derivation of the second-order correction term\cite{ExplicitA2for1980paper, SOPPA1980} is obtained by setting $n$ = 2 in the general expression for the $\mathbf{A}$ matrix given in Eq. (\ref{A.n'th.order}). In this way, an identical expression is recovered for both the singlet and triplet blocks of the second-order $\mathbf{A}$ matrix:
\beqa
A^{S(2)}_{ai,bj} = A^{T(2)}_{ai,bj} &=& 
\frac {1}{2} (\epsilon_b - \epsilon_j)
             \left ( \Del{ab} ~\rho^{(2)}_{ij}
                   - \Del{ij} ~\rho^{(2)}_{aa} \right ) 
\nonumber \\
 &&  - \frac{1}{2}\Del{ab}\sum_{cdk} \binp{jc}{kd} {}^{(1)}\mptcoef{cd}{ik}
 - \frac {1}{2} \Del{ij}\sum_{ckl} \binp{kb}{lc} {}^{(1)}\mptcoef{ac}{kl} .
\label{A.2nd.order}
\eeqa
The last two terms in Eq. (\ref{A.2nd.order}) are the standard expression for $\mathbf{A}^{(2)}$ given in the SOPPA formulations by Oddershede et. al (1977)\cite{Oddershede_1977_particle-hole} and Oddershede et. al (1984) \cite{ODDERSHEDE1984}. The first term corresponds to the missing term introduced by Nielsen et. al (1980)\cite{SOPPA1980}. This confirms that our approach, using the normalized RDMs to derive the corrections to the $\mathbf{A}$ matrix is valid.
%
It should be noted, that the $\mathbf{A}^{(2)}$ matrix needs to be made hermitian, so that for a real wavefunction a symmetrization must be carried out.\cite{ODDERSHEDE1984}

\subsubsection{The Third-Order Correction to the A matrix}
Setting $n=3$ in Eq. (\ref{A.n'th.order}) allows one to derive the complete third-order $\mathbf{A}$ matrix in terms of RDMs
\beqa
A^{X(3)}_{ai,bj} &=&
\frac {1}{2} (\epsilon_b - \epsilon_j)
             \left ( \Del{ab} ~\rho^{(3)}_{ij}
                   - \Del{ij} ~\rho^{(3)}_{ba} \right ) 
\nonumber \\
&\phantom{=}&- \left ( \sum _k \binp{ab}{kk}
- \frac{1}{4} \sum _k \binp{ak}{kb}
+ \frac{1}{4} \sum _c \binp{ac}{cb} \right )
\rho_{ij}^{(2)}\nonumber \\
&\phantom{=}&- \left ( \sum _k \binp{ji}{kk}
- \frac{1}{4} \sum _k \binp{jk}{ki}
+ \frac{1}{4} \sum _c \binp{jc}{ci} \right )
\rho_{ba}^{(2)}\nonumber \\
&\phantom{=}&+ ~\Del{ij} \sum _l \left ( \sum _k \binp{lb}{kk}
- \frac{1}{4} \sum _k \binp{lk}{kb}
+ \frac{1}{4} \sum _c \binp{lc}{cb} \right )
\rho_{la}^{(2)}\nonumber \\
&\phantom{=}&+ ~\Del{ij} \sum _d \left ( \sum _k \binp{db}{kk}
- \frac{1}{4} \sum _k \binp{dk}{kb}
+ \frac{1}{4} \sum _c \binp{dc}{cb} \right )
\rho_{da}^{(2)}\nonumber \\
&\phantom{=}&+ ~\Del{ab} \sum _l \left ( \sum _k \binp{jl}{kk}
- \frac{1}{4} \sum _k \binp{jk}{kl}
+ \frac{1}{4} \sum _c \binp{jc}{cl} \right )
\rho_{il}^{(2)}\nonumber \\
&\phantom{=}&+ ~\Del{ab} \sum _d \left ( \sum _k \binp{jd}{kk}
- \frac{1}{4} \sum _k \binp{jk}{kd}
+ \frac{1}{4} \sum _c \binp{jc}{cd} \right )
\rho_{id}^{(2)}\nonumber \\
&\phantom{=}& + \frac{1}{4} \sum _{pq} \bigg \{
\binp{jp}{qi} \left(  \pi_{bpqa}^{X(2)} + \pi_{qabp}^{X(2)} \right) +
\binp{ji}{pq} \left( \pi_{bapq}^{S(2)} + \pi_{pqba}^{S(2)} \right)
\nonumber \\
&\phantom{=}& \phantom{+ \frac{1}{4} \sum _{pq}} +
\binp{ap}{qb} \left( \pi_{ipqj}^{X(2)} +  \pi_{qjip}^{X(2)} \right)
+ \binp{ab}{pq} \left( \pi_{ijpq}^{S(2)} + \pi_{pqij}^{S(2)} \right)
\nonumber \\
&\phantom{=}& \phantom{+ \frac{1}{4} \sum _{pq}} - 
\binp{ap}{jq} \left( \pi_{ipbq}^{X(2)} + \pi_{bqip}^{X(2)} \right)
- \binp{pi}{qb} \left( \pi_{paqj}^{X(2)} +  \pi_{qjpa}^{X(2)} \right)
\bigg  \} \nonumber \\
&\phantom{=}& - \frac{1}{4} \Del{ij} \sum _{pqr}
\binp{pb}{qr} \left( \pi_{paqr}^{S(2)} +\pi_{qrpa}^{S(2)}
\right )
- \frac{1}{4} \Del{ab} \sum _{pqr}
\binp{jp}{qr}  \left ( \pi_{ipqr}^{S(2)} + \pi_{qrip}^{S(2)}
\right ) .
\label{A.3.order-1}
\eeqa
The full expression for the third- order contribution to the singlet \textbf{A} matrix, where the explicit expressions for the RDMs have been inserted can be found in appendix 1. Our derivation reproduces the three contributions, $\mathbf{A}^{(3)'}$, $\mathbf{A}^{(3)''}$ and $\mathbf{A}^{(3)'''}$, in the 1991 formulation\cite{TOPPA1991} of the singlet spin-adapted third-order $\mathbf{A}$ matrix apart from a few typos in the earlier presentation. However, we obtain an additional term, $\mathbf{A}^{(3)''''}$, not present in the 1977 \cite{Oddershede_1977_particle-hole} or 1991 formulations \cite{TOPPA1991}, containing the orbital energies
\beqa
A^{(3)''''}_{ai,bj} = \frac {1}{2} (\epsilon_b - \epsilon_j)
             \left ( \Del{ab} ~\rho^{(3)}_{ij}
                   - \Del{ij} ~\rho^{(3)}_{ba} \right ) . \label{eq:A3''''}
\eeqa
%
In the following, the SOPPA method augmented by the complete $\mathbf{A}^{(3)}$ matrix will be denoted as the SOPPA-A3 method, while SOPPA augmented by the incomplete $\mathbf{A}^{(3)}$ matrix, which is missing the contribution in \cref{eq:A3''''}, will be referred to as SOPPA-iA3 corresponding to the method presented by Eriksen and co-workers.\cite{Eriksen,TOPPA1991} 
A full implementation of the $\mathbf{A}^{(3)}$ matrix will scale as $N^6$ due to some of the terms involving the second-order 2RDMs in $\mathbf{A}^{(3)'}$ similar to the $\mathbf{B}^{(2)}$ matrix in SOPPA.
However, an efficient implementation of SOPPA-A3 would employ a Davidson like algorithm,\cite{SOPPAImplementationInDALTON1996, spas037} which avoids the explicit calculation of the matrices by directly calculating linear transformed trial vectors.
This will reduce the scaling again to $N^5$.

\section{Computational Details}

The performance of the SOPPA-A3 method was assessed
using two benchmarking sets for the calculation of vertical singlet excitation energies. 

The first set, called the Christiansen-Koch-Larsen set in the following, comprises 27 excited states of the five small molecules BH, CH$_2$, H$_2$O, HF, N$_2$ and the Ne atom and it was used in the calculations by Christiansen, Koch and Larsen. \cite{CHRISTIANSEN199847,KOCH1995,CHRISTIANSEN1996,Larsen} The basis sets and geometries used here are the same as in the aforementioned studies. For N$_2$ the cc-pVDZ basis was used \cite{dunning1989a} and for Ne, CH$_2$ and H$_2$O it was augmented with additional diffuse functions\cite{kendall1992a}. For HF the basis used was aug-cc-pVDZ\cite{dunning1989a,kendall1992a} and for BH it was the double augmented basis set d-aug-cc-pVDZ\cite{dunning1989a,kendall1992a,woon1994a}. Moreover, in the excitation energy calculations on H$_2$O, HF and N$_2$ the core orbitals were frozen.


The second set consists of 25 excited states of eight small molecules, BH, BF, CO, N$_2$, NH$_3$, H$_2$O, H$_2$S and HCl, taken from the benchmark studies of Loos \textit{et al.}\ \cite{Loos2018,Loos2021}, here called the Loos set.
The geometries and basis sets used are the same as in the aforementioned studies and the geometries are given in the Supplementary Information of reference. \cite{Loos2018} The basis set used for all molecules was the aug-cc-pVDZ basis set \cite{woon1993a,kendall1992a,dunning1989a}, and in the excitation energy calculations on NH$_3$, H$_2$O, CO, N$_2$, HCl and H$_2$S, the core orbitals were frozen.


The SOPPA, SOPPA-iA3 and SOPPA-A3 calculations were carried out in our in-house program (github.com/Linaknu/RPA-and-SOPPA-programs). 
Basis sets were taken from the Basis Set Exchange library. \cite{schuchardt2007a,feller1996a,pritchard2019a} and the calculations were performed with a residual tolerance of $10^{-4}$ au.

The excited states were matched across methods by comparing the dominant single-excitation amplitudes in the transition vectors.

\section{Results}

\subsection{The Christiansen-Koch-Larsen Set}
\begin{figure}[H]
\includegraphics[height=8cm]
{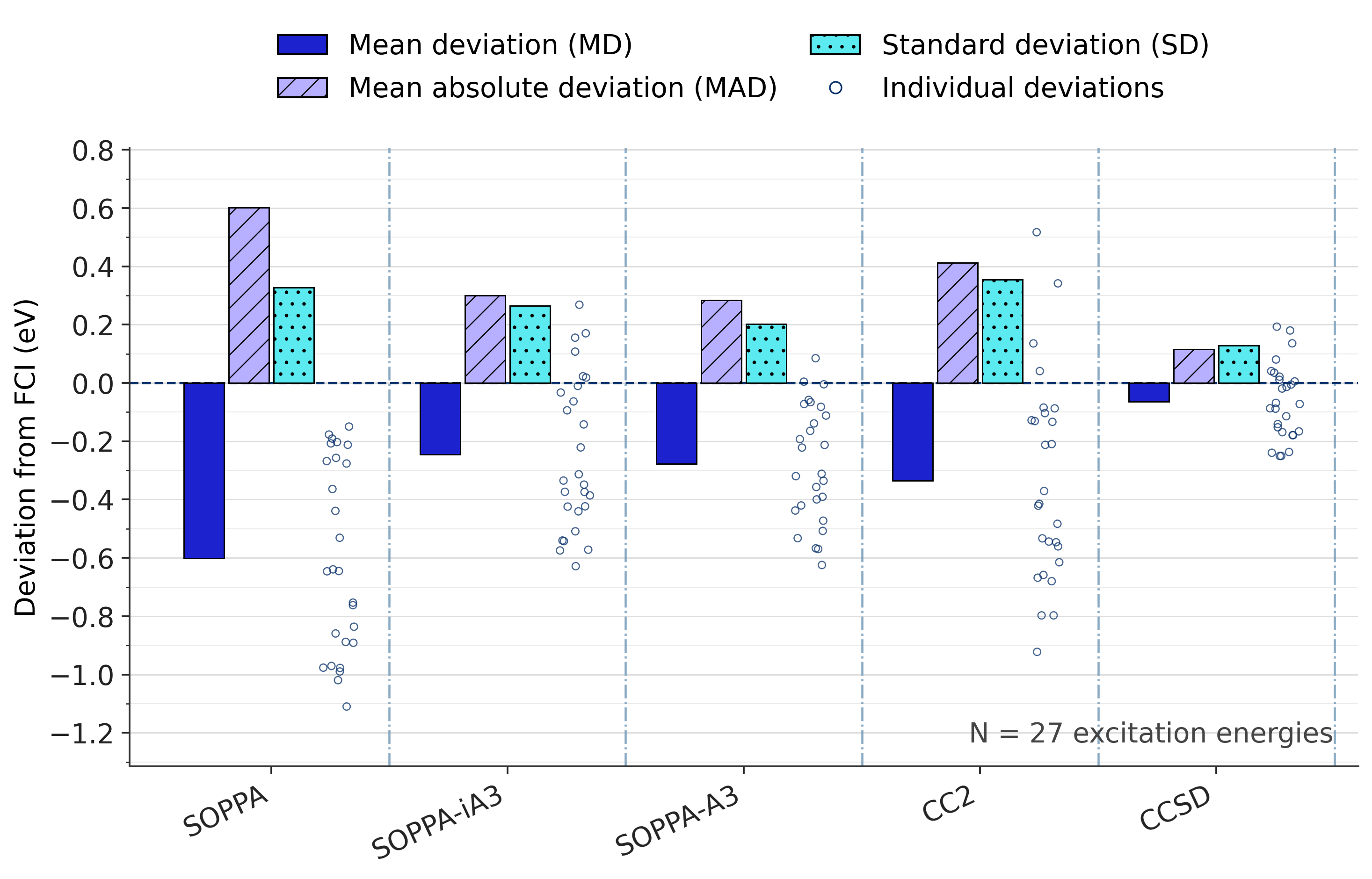}
\centering
\caption{Overview of the deviations of the singlet excitation energies from the FCI results for the SOPPA, SOPPA-iA3, SOPPA-A3, CC2 and CCSD methods for the Christiansen-Koch-Larsen set.}
\label{fig:FullFCI}
\end{figure}

The SOPPA, SOPPA-iA3, SOPPA-A3, CC2 and CCSD results for the 27 excited states in the Christiansen-Koch-Larsen set \cite{CHRISTIANSEN199847,CHRISTIANSEN1996,KOCH1995,Larsen} are presented in \cref{tab:Christiansen} as deviations from the corresponding FCI results.
In addition, the mean deviation (MD), mean absolute deviation (MAD) and standard deviation (SD) for each method are listed. A visual representation of the deviations are presented in \cref{fig:FullFCI}.

\begin{table}[H]
\centering
\begin{threeparttable}
\caption{Singlet excitation energies (eV) for the Christiansen-Koch-Larsen set shown as deviation from the FCI energies.
    }
\label{tab:Christiansen}
\begin{tabular}{llr|rrrrr}
\hline
 Mol. & State & FCI\tnote{a} & SOPPA & SOPPA-iA3 & SOPPA-A3 & CC2\tnote{a} & CCSD\tnote{a}\\  
 \hline
 
\ce{Ne} & ${}^{1}P^{}$  & 16.398 & $-$0.859 & $-$0.440 & $-$0.438 & $-$0.659 & $-$0.239 \\ 
     & ${}^{1}D$     & 18.213 & $-$0.978 & $-$0.572 & $-$0.567 & $-$0.797 & $-$0.250 \\
     & ${}^{1}P$     & 18.256 & $-$0.977 & $-$0.575 & $-$0.570 & $-$0.797 & $-$0.250 \\
     & ${}^{1}S$     & 18.485 & $-$1.110 & $-$0.628 & $-$0.624 & $-$0.922 & $-$0.237 \\
     & ${}^{1}S$     & 44.053 & $-$0.531 & $-$0.063 & $-$0.072 & $-$0.371\tnote{b} & $-$0.169 \\[1.0ex]

  \ce{BH}  & ${}^{1}\Pi$        & 2.944 & $-$0.439 & 0.170 & $-$0.066 & $-$0.084 & 0.022 \\
       & ${}^{1}\Sigma^{+}$  & 6.378 & $-$0.177 & 0.155 & $-$0.004 & $-$0.087 & 0.041 \\
       & ${}^{1}\Pi$ & 7.467 & $-$0.207 & 0.108 & $-$0.058 & $-$0.133 & 0.035 \\
       & ${}^{1}\Sigma^{+}$  & 7.559 & $-$0.276 & 0.023 & $-$0.139 & $-$0.212 & 0.193 \\[1.0ex]

 \ce{H2O} & ${}^{1}B_{1}$  & 7.447 & $-$0.645 & $-$0.424 & $-$0.420 & $-$0.414 & $-$0.072 \\
 &${}^{1}A_{2}$            & 9.211 & $-$0.753 & $-$0.542 & $-$0.532 & $-$0.546 & $-$0.088 \\ 
 &${}^{1}A_{1}$           & 9.874 & $-$0.646 & $-$0.386 & $-$0.390 & $-$0.421 & $-$0.068 \\
 &${}^{1}B_{2}$            & 11.612 & $-$0.762 & $-$0.509 & $-$0.507 & $-$0.561 & $-$0.087 \\[1.0ex]

  \ce{CH2} & ${}^{1}B_{2}$  & 1.793 & $-$0.639 & $-$0.033 & $-$0.222 & $-$0.128 & $-$0.014 \\
           & ${}^{1}A_{2}$  & 5.853 & $-$0.212 & 0.269 & 0.085 & 0.041 & 0.005 \\
           & ${}^{1}A_{1}$  & 6.514 & $-$0.191 & 0.019 & $-$0.082 & $-$0.103 & $-$0.005 \\
           & ${}^{1}B_{1}$  & 7.704 & $-$0.203 & $-$0.010 & $-$0.112 & $-$0.131 & 0.011 \\
           & ${}^{1}A_{1}$  & 8.479 & $-$0.268 & $-$0.094 & $-$0.192 & $-$0.210 & $-$0.019 \\[1.0ex]

 \ce{HF} & ${}^{1}\Pi$  &  10.440 & $-$0.971 & $-$0.335 & $-$0.320 & $-$0.615 & $-$0.141 \\
      & ${}^{1}\Pi$  &  14.214 & $-$1.020 & $-$0.423 & $-$0.399 & $-$0.680 & $-$0.152 \\
      & ${}^{1}\Sigma^{+}$  &  14.582 & $-$0.836 & $-$0.221 & $-$0.213 & $-$0.483 & $-$0.114 \\
      & ${}^{1}\Delta$  &  15.197 & $-$0.891 & $-$0.349 & $-$0.312 & $-$0.533 & $-$0.166 \\
      & $1 {}^{1}\Sigma^{-}$  &  15.278 & $-$0.888 & $-$0.373 & $-$0.336 & $-$0.544 & $-$0.179 \\
      & ${}^{1}\Pi$  &  15.771 & $-$0.990 & $-$0.374 & $-$0.357 & $-$0.668 & $-$0.179 \\[1.0ex]

  \ce{N2}  & ${}^{1}\Pi_{g}$         & 9.584 & $-$0.364 & $-$0.540 & $-$0.472 & 0.136 & 0.081 \\
          & ${}^{1}\Sigma_{u}^{-}$  & 10.329 & $-$0.257 & $-$0.313 & $-$0.164 & 0.342 & 0.136 \\
          & ${}^{1}\Delta_{u}$      & 10.718 & $-$0.150 & $-$0.142 & 0.005 & 0.517 & 0.180 \\[1.0ex]
  \hline
  \multicolumn{3}{l} {Mean Absolute Deviation}   & 0.580 & 0.289 & 0.273 & 0.398 & 0.112 \\

  \multicolumn{3}{l} {Mean Deviation} & $-$0.601 & $-$0.245 & $-$0.277 & $-$0.336 & $-$0.064 \\

   \multicolumn{3}{l} {Minimum}  & $-$1.110 & $-$0.628 & $-$0.624 & $-$0.922 & $-$0.250 \\

  \multicolumn{3}{l} {Maximum} & $-$ & 0.269 & 0.085 & 0.517 & 0.193 \\

   \multicolumn{3}{l} {Standard Deviation}   & 0.327 & 0.264 & 0.202 & 0.355 & 0.128 \\
\hline
\end{tabular}
\begin{tablenotes}
\item[a] H$_2$O and N$_2$ from reference \cite{CHRISTIANSEN1996}. Ne, CH$_2$, and BH from reference \cite{KOCH1995}. HF from reference \cite{Larsen}. 
\item[b] replaced by the corrected value from the 1998 publication \cite{CHRISTIANSEN199847}
\end{tablenotes}
\end{threeparttable}
\end{table}

\cref{fig:FullFCI} clearly shows 
that all methods tend to underestimate the excitation energies and that CCSD outperforms the other methods on all parameters.

SOPPA-A3 achieves a smaller MAD and SD than both SOPPA and CC2. The improvement of SOPPA-A3 over SOPPA is substantial: with the inclusion of the third order contribution to the \textbf{A} matrix, the MAD is halved from $0.580$ eV to $0.273$ eV, while the SD is reduced by approximately a third from $0.327$ eV to $0.202$ eV. The SOPPA-iA3 and SOPPA-A3 methods are found to have a MAD of $0.289$ eV and $0.273$ eV, respectively. However, SOPPA-iA3 yields a higher SD of $0.264$ eV compared to SOPPA-A3 with $0.202$ eV. 

SOPPA-A3 performs better than SOPPA for all excitation energies except for the 1${}^{1}\Pi_{g}$ transition in N$_{2}$. The largest improvements are generally found for Ne and HF, for which SOPPA performs poorly. The excitation energy for Ne that shows the largest change increases by $0.486$ eV for SOPPA-A3 compared with SOPPA for which the deviations from FCI are $-0.624$ eV and $-1.110$ eV, respectively. For HF the largest change is $0.651$ eV for the excitation energy determined to have a deviation from FCI of $-0.320$ eV and $-0.971$ eV for SOPPA-A3 and SOPPA, respectively.

In comparison to the MAD of CC2, $0.398$ eV, the MAD of SOPPA-A3, $0.273$ eV, is smaller by approximately a fourth.
Furthermore, the SD of SOPPA-A3, $0.202$ eV, is smaller than for CC2, $0.355$ eV, by more than a third.
For H$_{2}$O the deviations from the FCI results of the SOPPA-A3 and the CC2 methods are very similar, while for Ne, HF and BH, the SOPPA-A3 method is found to perform consistently better than CC2. 
As observed for SOPPA-A3 relative to SOPPA, the improvement is most pronounced for Ne and HF. If Ne and HF were not included, the MAD of SOPPA-A3 would only be $0.079$ eV lower than for CC2.

\subsection{The Loos Set}
\begin{figure}[H]
\includegraphics[height=8cm]
{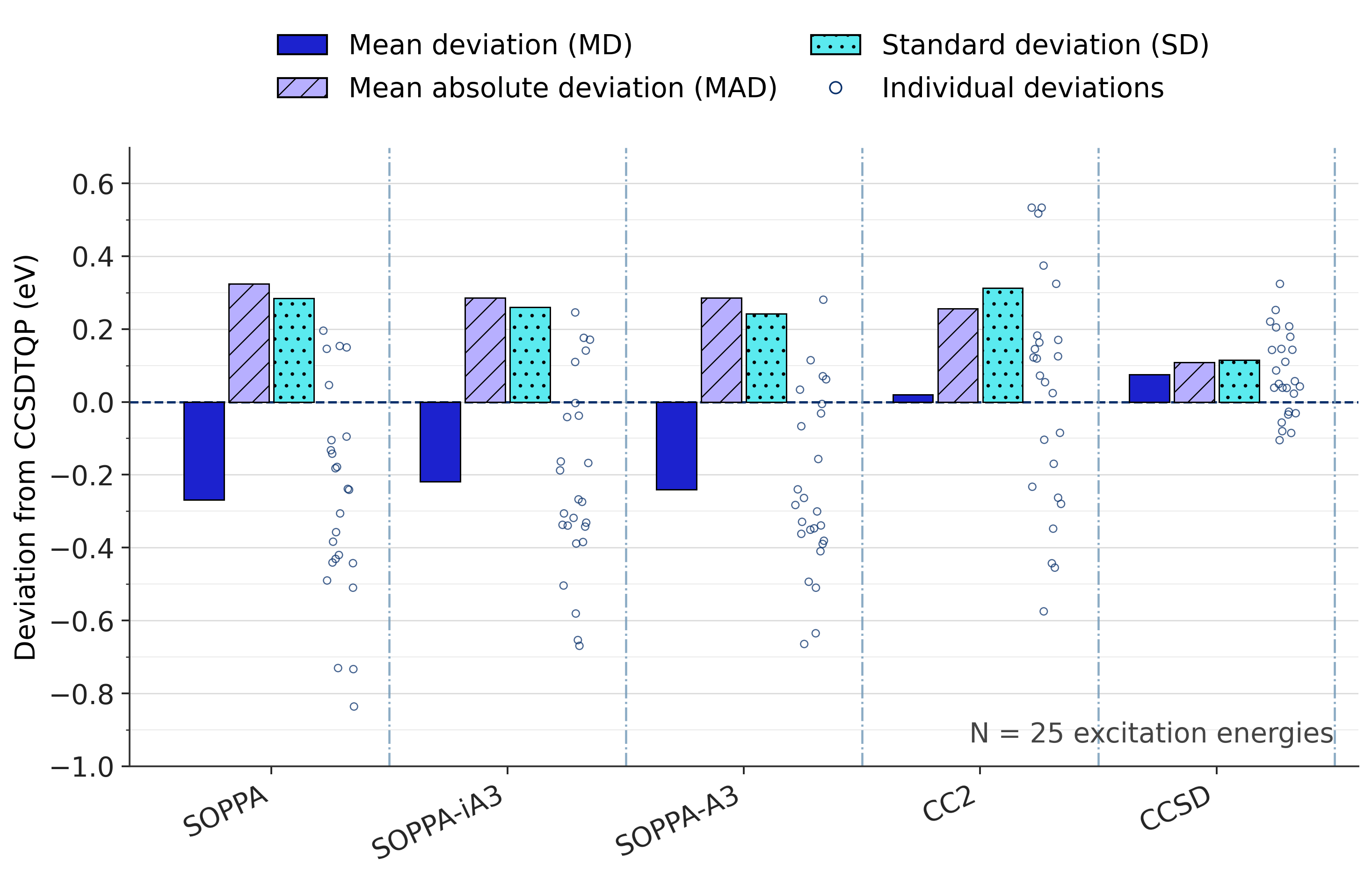}
\centering
\caption{Overview of the deviations of the singlet excitation energies to the CCSDTQP results of the SOPPA, SOPPA-iA3, SOPPA-A3, CC2 and CCSD methods for the Loos set.}
\label{fig:Loos}
\end{figure}

The SOPPA, SOPPA-iA3 and SOPPA-A3 results for the 25 excited states in the Loos set\cite{Loos2018,Loos2021} are presented in \cref{tab:Loos}, where the excitation energies are shown as deviations from the corresponding CCSDTQP results.
For comparison, deviations of corresponding CC2 and CCSD results from the CCSDTQP values are included. Visual representations of the deviations are presented in \cref{fig:Loos}.

\begin{table}[H]
\begin{threeparttable}
\caption{Singlet excitation energies (eV) for the Loos set shown as differences from the CCSDTQP energies.
  }
\label{tab:Loos}
\begin{tabular}{llr|rrrrr}
\hline
Mol. & State & CCSDTQP\tnote{a} & SOPPA & SOPPA-iA3 & SOPPA-A3 & CC2\tnote{a} & CCSD\tnote{a}\\  
\hline

  \ce{BH} &${}^{1}\Pi$ & 2.947 &$-$0.441& 0.171 &$-$0.067& $-$0.085 & 0.023 \\[1.0ex]

\ce{NH3} & ${}^{1}A_{2}$  & 6.482 & $-$0.420 & $-$0.274 & $-$0.301 & $-$0.233 & $-$0.027 \\
& ${}^{1}E$    & 8.081  & $-$0.510 & $-$0.389 & $-$0.410 & $-$0.348 & $-$0.057  \\
& ${}^{1}A_1$  & 9.680  & $-$0.490 & $-$0.340 & $-$0.362 & $-$0.280 & $-$0.031\\
& ${}^{1}A_2$  & 10.411 & $-$0.443 & $-$0.306 & $-$0.339 & $-$0.263 & $-$0.035\\[1.0ex]

  \ce{H2O} &${}^{1}B_{1}$ & 7.532 & $-$0.731 & $-$0.385 & $-$0.381 & $-$0.443 & $-$0.085 \\
           &${}^{1}A_{2}$ & 9.318 & $-$0.836 & $-$0.504 & $-$0.494 & $-$0.575 & $-$0.105 \\
           &${}^{1}A_{1}$ & 9.941 & $-$0.734 & $-$0.342 & $-$0.347 & $-$0.455 & $-$0.080 \\

  \ce{H2S} &${}^{1}B_{1}$ & 6.103 & $-$0.095 & 0.109 & 0.034 & 0.054 & 0.038\\
           &${}^{1}A_{2}$ & 6.286 & $-$0.105 & 0.245 & 0.114 & 0.145 & 0.057\\[1.0ex]
  
  \ce{HCl} & ${}^{1}\Pi$   & 7.823 & $-$0.133 & 0.141 & 0.062 & 0.072 & 0.039 \\[1.0ex]

  \ce{BF} &${}^{1}\Pi$ & 6.485 & $-$0.358 & $-$0.003 & $-$0.240 & 0.024 & 0.049\\[1.0ex]

 \ce{CO} &${}^{1}\Pi$         & 8.561  & $-$0.241 & $-$0.337 & $-$0.390 & 0.163 & 0.110 \\
      &${}^{1}\Sigma^{-}$  & 10.057 & $-$0.431 & $-$0.038 & 0.070 & 0.324 & 0.039  \\
      &${}^{1}\Delta$      & 10.168 & $-$0.306 & 0.176 & 0.280 & 0.517 & 0.042 \\
      &${}^{1}\Sigma^{+}$  & 10.919 & 0.195 & $-$0.268 & $-$0.351 & 0.170 & 0.252 \\
      &${}^{1}\Sigma^{+}$  & 11.506 & 0.149 & $-$0.168 & $-$ 0.264 & 0.122 & 0.204 \\
      &${}^{1}\Pi$         & 11.753 & 0.153 & $-$0.188 & $-$0.283 & 0.125 & 0.220 \\[1.0ex]

  \ce{N2} & ${}^{1}\Pi_{g}$         &  9.409  & $-$0.384 & $-$0.581 & $-$0.510 & 0.119 & 0.086 \\
          & ${}^{1}\Sigma_{u}^{-}$  & 10.054 & $-$0.239 & $-$0.318 & $-$0.157 & 0.374 & 0.143\\
          & ${}^{1}\Delta_{u}$      & 10.428 & $-$0.142 & $-$0.164 & $-$0.006 & 0.533 & 0.179 \\
          & ${}^{1}\Sigma_{u}^{+}$  & 13.107 & $-$0.179 & $-$0.670 &  $-$0.655 & $-$0.170 & 0.142 \\
          &  ${}^{1}\Pi_{u}$        & 13.127 & 0.146 & $-$0.332 & $-$0.329 & 0.182 & 0.324 \\
          & ${}^{1}\Sigma_{g}^{+}$  & 13.181 & $-$0.182 & $-$0.654 & $-$0.635 & $-$0.104 & 0.145 \\
          & ${}^{1}\Pi_{u}$         & 13.558 & 0.046 & $-$0.042 & $-$0.032 & 0.533 & 0.207 \\[1.0ex]

  \hline  
  \multicolumn{3}{l} {Mean Absolute Deviation} & 0.324 & 0.286 & 0.285 & 0.257 &  0.109 \\

  \multicolumn{3}{l} {Mean Deviation}   & $-$0.268 & $-$0.218 & $-$0.240 &  0.020  & 0.075 \\ 

  \multicolumn{3}{l} {Minimum}   & $-$0.836 & $-$0.670 & $-$0.655 & $-$0.575 & $-$0.105\\

  \multicolumn{3}{l} {Maximum}   & 0.195 & 0.245 & 0.280 & 0.533 &  0.324 \\ 
  
  \multicolumn{3}{l} {Standard Deviation}  & 0.284 & 0.260 & 0.242 & 0.312 &  0.115\\
\hline
\end{tabular}
\begin{tablenotes}
    \item[a] from reference \cite{Loos2021}.
\end{tablenotes}
\end{threeparttable}
\end{table}


As expected, CCSD was found to perform the best with the smallest SD and MAD of 0.115 and 0.109, respectively. The CC2 and CCSD methods tend to overestimate the excitation energies, while SOPPA, SOPPA-iA3 and SOPPA-A3 all tend to underestimate the excitation energies. 
SOPPA-A3 produces a slightly higher MAD of $0.285$ eV compared to the $0.257$ eV of CC2, but a lower SD of $0.241$ eV compared to the $0.312$ eV of CC2.
%
%
%
%
SOPPA-A3 with a MAD of $0.285$ eV is generally an improvement over SOPPA, with and MAD of $0.324$ eV. 
The improvement is largely due to the SOPPA outliers with a deviation from FCI around $-0.8$ eV in \cref{fig:Loos}, which are the three states belonging to H$_{2}$O. 
These states are more "well-behaved" for SOPPA-A3. 

The SOPPA-iA3 and SOPPA-A3 methods yield a very similar MAD of 0.286 eV and 0.285 eV, respectively, and a SD of 0.260 eV and 0.242 eV, respectively. 
In \cref{fig:Loos} and \cref{tab:Loos}, the difference between the two methods is found to be small for most states, with the largest differences being -$0.238$ eV for the ${}^{1}\Pi$ state of BH, and -$0.237$ eV for the ${}^{1}\Pi$ state of BF.
The effect of including the fourth contribution is therefore evidently state dependent; Some excitation energies are lowered while others are increased, even for transitions in the same molecule, e.g. CO and $\mathrm{N_2}$. 

\section{Conclusion}

We have derived a general expression for the spin-adapted $n$'th-order correction to the $\mathbf{A}$ matrix expressed in terms of one- and two-electron reduced density matrices.
The general formulation provides a unified framework for constructing higher-order corrections and enabled a consistent re-derivation of the third-order $\mathbf{A}$ matrix, 
which has allowed us to identify a contribution that has been omitted in earlier derivations and to correct errors in the remaining terms. 
We have obtained the complete singlet and triplet third-order $\mathbf{A}$ matrices and as a result report the first calculations of singlet excitation energies at the level of the Second-Order Polarization Propagator Approximation augmented with the complete third-order $\mathbf{A}$ matrix.

The method was tested on two benchmark sets comprising ten small molecules and one atom. For the Christiansen-Koch-Larsen benchmark set, the inclusion of the complete third-order correction leads to a substantial improvement over standard SOPPA. SOPPA-A3 outperforms both SOPPA and CC2 with respect to both accuracy and precision and yields lower deviations from FCI results than SOPPA for all but a single transition. 
A somewhat different picture emerges for the Loos benchmark set. Here, SOPPA-A3 remains consistently more accurate than SOPPA, but CC2 exhibits a slightly lower mean absolute deviation. 

The comparison between SOPPA-A3 and SOPPA-iA3 provides insight into the significance of the newly identified fourth contribution to the third-order $\mathbf{A}$ matrix. Across both benchmark sets, the two methods exhibit very similar mean absolute deviations and standard deviations, indicating that the dominant third-order effects were already captured by SOPPA-iA3. However, the effect of the newly included contribution is transition dependent both with regards to the size and direction of the deviation from CCSDTQP. 
We conclude that the fourth contribution cannot be considered universally insignificant and should be included for a formally complete third-order treatment.

The developments in this work provide a solid foundation for future investigations of higher-order polarization propagator approximations and their application to the calculation of electronically excited states.
Furthermore, it represents the first step towards the calculation of linear response properties such as frequency-dependent polarizabilities or the spin-spin coupling constants of nuclear magnetic resonance spectroscopy.


\clearpage

\section{Appendix 1}
The equation for the $\mathbf{A}^{(3)}$ contribution expressed in 2-electron integrals and MP correlation coefficients. The $\mathbf{A}^{(3)}$ matrix has conventionally been written as a sum of three terms. Including the missing term as the fourth term in the expression, the matrix takes the form
\begin{equation}
    \mathbf{A}^{(3)} = \mathbf{A}^{(3)'} + \mathbf{A}^{(3)''} + \mathbf{A}^{(3)'''} +  \mathbf{A}^{(3)''''} .
    \label{Contributions_A3}
\end{equation}
$\mathbf{A}^{(3)'''}$ includes the terms that contain the OVOV and VOVO blocks of the second-order two-particle RDM,  \cref{eq.pi_iajb,eq.pi_iajb-t}, and thus the second-order doubles correlation coefficients $^{(2)}\Kapg{ab}{ij}$. 
$\mathbf{A}^{(3)''}$ includes all the contributions containing the OV or VO blocks of the second-order one-particle RDM, \cref{D2ai}, i.e. the second-order singles correlation coefficients $^{(2)}\Kapg{a}{i}$. $\mathbf{A}^{(3)'}$ includes all the remaining contributions, which involve the OO or VV blocks of the second-order one-particle RDMs, \cref{D2ab,D2ij}, or other products of first-order doubles correlation coefficients $^{(1)}\Kapg{ab}{ij}$. $\mathbf{A}^{(3)''''}$ includes the contributions from the third-order one-particle RDMs.
In the following expression, the singlet spin-adapted doubles amplitudes\cite{mbpt770701p} are employed: 
\beqa
\Kap{1} &=& 
\left( \Kapg{ab}{ij} + \Kapg{ab}{ji}\right) \\
\Kap{2} &=& 
\left( \Kapg{ab}{ij} - \Kapg{ab}{ji}\right)
\eeqa

%
\begin{equation} 
    \begin{split}
A&^{(3)'}_{a i,b j} = 
\\ & - \delta_{ab} \bigg[ \frac{1}{4}
    \sum_{\substack{c d\\k l m}} \left\{ (j m| k i) - 2(j i| km) \right\}  \left\{  {}^{(1)}\kappa_{ kl}^{cd}(1)  {}^{(1)}\kappa_{m l}^{cd}(1) + {}^{(1)}\kappa_{ k l}^{cd}(2) {}^{(1)}\kappa_{ m l}^{cd}(2)
    \right\} 
    \\ 
    &\qquad\quad + \frac{1}{4} 
    \sum_{\substack{c d e\\k l}} 
    \left\{ 2 (j i|e c) - (j c|e i)
    \right\}\left\{
        {}^{(1)}\kappa_{kl}^{cd}(1)
        {}^{(1)}\kappa_{kl}^{ed}(1)
        +
        {}^{(1)}\kappa_{kl}^{cd}(2)
        {}^{(1)}\kappa_{kl}^{ed}(2)
    \right\}
    \\ 
    &\qquad\quad + \frac{1}{4}
    \sum_{\substack{c d\\k l m }}
    (j l| m k)
    \left\{
        {}^{(1)}\kappa_{m i}^{cd}(1)
        {}^{(1)}\kappa_{k l}^{cd}(1)
        +
        {}^{(1)}\kappa_{m i}^{cd}(2)
        {}^{(1)}\kappa_{k l}^{cd}(2)
    \right\}
    \\ 
    &\qquad\quad - \frac{1}{2}
    \sum_{\substack{c d e\\k l}} 
    (j l|c e)
    \left\{
        {}^{(1)}\kappa_{k i}^{ed}(1)
        {}^{(1)}\kappa_{k l}^{cd}(1)
        +
        {}^{(1)}\kappa_{k i}^{ed}(2)
        {}^{(1)}\kappa_{k l}^{cd}(2)
    \right\}
    \\ 
    &\qquad\quad + \frac{1}{4}
    \sum_{\substack{c d e\\k l}} 
    (j e|c k)
    \left\{
        {}^{(1)}\kappa_{i l}^{ed}(1)
        +
        \sqrt{3}{{}^{(1)}\kappa_{i l}^{ed}(2)}
    \right\}
    \left\{
        {}^{(1)}\kappa_{k l}^{cd}(1)
        +
        \sqrt{3}{{}^{(1)}\kappa_{k l}^{cd}(2)}
    \right\}
    \bigg] 
    \\ 
    &- \delta_{i j}
    \bigg[ \frac{1}{4}
        \sum_{\substack{c d e\\k l}} 
        \left\{
            (e b|a c) - 2(a b|e c)
        \right\}
        \left\{
            {}^{(1)}\kappa_{k l}^{cd}(1)
            {}^{(1)}\kappa_{k l}^{ed}(1)
            +
            {}^{(1)}\kappa_{k l}^{cd}(2)
            {}^{(1)}\kappa_{k l}^{ed}(2)            
        \right\}
        \\ 
    &\qquad\quad + \frac{1}{4}
    \sum_{\substack{c d\\k l m}} 
    \left\{
        2 (a b|k m) - (k b|a m)
    \right\}
    \left\{
        {}^{(1)}\kappa_{k l}^{cd}(1)
        {}^{(1)}\kappa_{m l}^{cd}(1)
        +
        {}^{(1)}\kappa_{k l}^{cd}(2)
        {}^{(1)}\kappa_{m l}^{cd}(2)
    \right\}
    \\ 
    &\qquad\quad + \frac{1}{4}
    \sum_{\substack{c d e\\k l}} 
    (d b|c e)
    \left\{
        {}^{(1)}\kappa_{k l}^{ea}(1)
        {}^{(1)}\kappa_{k l}^{cd}(1)
        +
        {}^{(1)}\kappa_{k l}^{ea}(2)
        {}^{(1)}\kappa_{k l}^{cd}(2) 
    \right\}
    \\ 
    &\qquad\quad - \frac{1}{2} 
    \sum_{\substack{c d\\k l m}} 
    (d b| m k)
    \left\{
        {}^{(1)}\kappa_{m l}^{ca}(1)
        {}^{(1)}\kappa_{k l}^{cd}(1)
        +
        {}^{(1)}\kappa_{m l}^{ca}(2)
        {}^{(1)}\kappa_{k l}^{cd}(2)         
    \right\}
    \\ 
    &\qquad\quad + \frac{1}{4} 
    \sum_{\substack{c d\\k l m}} 
    (m b|c k)  
    \left\{
        {}^{(1)}\kappa_{m l}^{ad}(1)
        +
        \sqrt{3} {}^{(1)}\kappa_{m l}^{ad}(2)
    \right\}
    \left\{
        {}^{(1)}\kappa_{k l}^{cd}(1)
        +
        \sqrt{3} {}^{(1)}\kappa_{k l}^{cd}(2)
    \right\}
    \bigg]
    \\ 
    & + \frac{1}{4}
    \sum_{\substack{c d\\k l}} 
    \left\{(ji|dc) + (jc|di)\right\}
    {}^{(1)}\kappa_{kl}^{ac}(1)
    {}^{(1)}\kappa_{kl}^{bd}(1)
    \\ 
    & + \frac{1}{4}
    \sum_{\substack{c d\\k l}} 
    \left\{(ji|dc) - (jc|di)\right\}
    {}^{(1)}\kappa_{kl}^{ac}(2)
    {}^{(1)}\kappa_{kl}^{bd}(2)
    \\ 
    &- \frac{1}{2}
    \sum_{\substack{c\\k l m}}  
    (j i| k m)
    \left\{
        {}^{(1)}\kappa_{k l}^{ac}(1)
        {}^{(1)}\kappa_{m l}^{bc}(1)
        +
        {}^{(1)}\kappa_{k l}^{ac}(2)
        {}^{(1)}\kappa_{m l}^{bc}(2)
    \right\}
    \nonumber 
\end{split}
\end{equation}
\begin{equation}
    \begin{split}
    \phantom{A'}&+ \frac{1}{4} 
    \sum_{\substack{c\\k l m}} 
    (j m| k i)
    \left\{
        {}^{(1)}\kappa_{k l}^{ac}(1)
        +
        \sqrt{3}   
        {}^{(1)}\kappa_{k l}^{ac}(2)
    \right\}
    \left\{
       {{}^{(1)}\kappa_{m l}^{bc}(1)} 
       +
       \sqrt{3}    
       {}^{(1)}\kappa_{m l}^{bc}(2)
    \right\}  
    \\ 
    &- \sum_{\substack{c d\\k l}} 
    \left\{
        \frac{1}{2}       
        (j d|a i) - \frac{1}{4}(j i|a d)
    \right\}
    \left\{
        {}^{(1)}\kappa_{k l}^{cd}(1)
        {}^{(1)}\kappa_{k l}^{cb}(1)
        +
        {}^{(1)}\kappa_{k l}^{cd}(2)
        {}^{(1)}\kappa_{k l}^{cb}(2)
    \right\}
    \\ 
    &- \sum_{\substack{c d\\k l}}  
    \left\{
        \frac{1}{2}   
        (a k| j b) - \frac{1}{4}(a b| j k)
    \right\}
    \left\{
        {}^{(1)}\kappa_{k l}^{cd}(1)
        {}^{(1)}\kappa_{i l}^{cd}(1)
        +
        {}^{(1)}\kappa_{k l}^{cd}(2)
        {}^{(1)}\kappa_{i l}^{cd}(2)
    \right\}
    \\ 
    &- \frac{1}{4} 
    \sum_{\substack{c d\\k l}}
    (j l|a d)
    \left\{
        {}^{(1)}\kappa_{k i}^{cd}(1)
        +
        \sqrt{3}    
        {}^{(1)}\kappa_{k i}^{cd}(2)
    \right\}
    \left\{
        {}^{(1)}\kappa_{l k}^{bc}(1)
        +
        \sqrt{3}    
        {}^{(1)}\kappa_{l k}^{bc}(2)
    \right\}
    \\ 
    &+ \frac{1}{2} 
    \sum_{\substack{c d\\k l}}  
    (j d|a l)
    \left\{
        {}^{(1)}\kappa_{k i}^{cd}(1)
        {}^{(1)}\kappa_{l k}^{bc}(1)
        +
        {}^{(1)}\kappa_{k i}^{cd}(2)
        {}^{(1)}\kappa_{l k}^{bc}(2)        
    \right\}
    \\ 
    &\ + \frac{1}{4}
    \sum_{\substack{c d\\k l}}
    \left\{(ab|lk) + (lb|ak)\right\}
    {}^{(1)}\kappa_{il}^{cd}(1)
    {}^{(1)}\kappa_{jk}^{cd}(1)
    \\ 
    &    + \frac{1}{4}
    \sum_{\substack{c d\\k l}}  
    \left\{(ab|lk) - (lb|ak)\right\}
    {}^{(1)}\kappa_{il}^{cd}(2)
    {}^{(1)}\kappa_{jk}^{cd}(2)       
    \\ 
    &- \frac{1}{2}
    \sum_{\substack{c d e\\k}}  
    (a b|e c)
    \left\{
        {}^{(1)}\kappa_{i k}^{cd}(1)
        {}^{(1)}\kappa_{j k}^{ed}(1)
        +
        {}^{(1)}\kappa_{i k}^{cd}(2)
        {}^{(1)}\kappa_{j k}^{ed}(2)       
    \right\}
    \\ 
    &+ \frac{1}{4}
    \sum_{\substack{c d e\\k}}      
    (e b| ac)
    \left\{
        {}^{(1)}\kappa_{i k}^{cd}(1)
        +
        \sqrt{3}    
        {}^{(1)}\kappa_{i k}^{cd}(2)
    \right\}
    \left\{
        {}^{(1)}\kappa_{j k}^{ed}(1)
        +
        \sqrt{3}    
       {{}^{(1)}\kappa_{j k}^{ed}(2)}      
    \right\}
    \\ 
    &- \sum_{\substack{c d\\k l}}    
    \left\{
        \frac{1}{2}   
        (l b|a i) -
        \frac{1}{4}
        (a b|l i)
    \right\}
    \left\{
        {}^{(1)}\kappa_{k l}^{cd}(1)
        {}^{(1)}\kappa_{k j}^{cd}(1)
        +
        {}^{(1)}\kappa_{k l}^{cd}(2)
        {}^{(1)}\kappa_{k j}^{cd}(2)        
    \right\}
    \\ 
    &- \sum_{\substack{c d\\k l}} 
    \left\{
        \frac{1}{2}   
        (c i| j b)
        -
        \frac{1}{4}
        (j i|cb)
    \right\}
    \left\{
        {}^{(1)}\kappa_{k l}^{cd}(1)
        {}^{(1)}\kappa_{k l}^{ad}(1)
        +
        {}^{(1)}\kappa_{k l}^{cd}(2)    
        {}^{(1)}\kappa_{k l}^{ad}(2)    
    \right\}
    \\ 
    &- \frac{1}{4}
    \sum_{\substack{c e\\k l}} 
    (e b|l i)
    \left\{
        {}^{(1)}\kappa_{k l}^{ca}(1)
        +
        \sqrt{3}    
        {}^{(1)}\kappa_{k l}^{ca}(2)
    \right\}  
    \left\{
        {}^{(1)}\kappa_{j k}^{ec}(1)
        +
        \sqrt{3}    
        {}^{(1)}\kappa_{j k}^{ec}(2)
    \right\}
    \\ 
    &+ \frac{1}{2}
    \sum_{\substack{c d\\k l}}  
    (l b|d i)
    \left\{
        {}^{(1)}\kappa_{k l}^{ca}(1)
        {}^{(1)}\kappa_{j k}^{dc}(1)
        +
        {}^{(1)}\kappa_{k l}^{ca}(2)
        {}^{(1)}\kappa_{j k}^{dc}(2)
    \right\}
\end{split}
\end{equation}

\begin{eqnarray}
    A^{(3)''}_{a i, b j} 
    &=\!\!\!\!& 
    - \frac{1}{2} \delta_{ab} 
    \sum_{ck} \left\{ 4(j i|c k) - (j k|c i) -(j c|k i) \right \} 
    ~\rho^{(2)}_{kc}
    \nonumber \\ 
    &&- \frac{1}{2} \delta_{i j}
    \sum_{ck} \left \{(a k|c b) + (a c|k b) - 4(a b|c k) \right \}
    ~\rho^{(2)}_{kc}
    \nonumber \\  
    &&- 
    \sum_ k \left\{  (j k|a i) ~\rho^{(2)}_{kb} 
    + (k i| j b) ~\rho^{(2)}_{ka} \right\} 
    - \sum_c \left\{ (a i|c b) ~\rho^{(2)}_{jc}
    + (a c|j b) ~\rho^{(2)}_{ic} \right\}
    \nonumber \\ 
    &&- \frac{1}{2} \sum_c \left\{ (a b|c i) ~\rho^{(2)}_{jc}
    + (a b|j c) ~\rho^{(2)}_{ic} \right \}
    - \frac{1}{2} \sum_ k  \left\{ (j i|a k) ~\rho^{(2)}_{kb} 
    + (j i| k b) ~\rho^{(2)}_{ka} \right \} \\
    &=& 
- \frac{\sqrt{2}}{2} \delta_{ab} \sum_{ck} \left[4(j i|c k) - (j k|c i) -(j c|k i) 
    \right] {}^{(2)}\kappa_ k^c
    \nonumber \\ 
    &&- \frac{\sqrt{2}}{2} \delta_{i j}
    \sum_{ck} \left[
         (a k|c b) + (a c|k b) - 4(a b|c k) 
    \right] {}^{(2)}\kappa_ k^c
    \nonumber \\  
    &&- \sqrt{2}
    \left[  \sum_ k \left\{  (j k|a i)         {}^{(2)}\kappa_ k^b + (k i| j b) {}^{(2)}\kappa_ k^a        \right\} - \sum_c \left\{ (a i|c b) {}^{(2)}\kappa_ j^c
    + (a c|j b){}^{(2)}\kappa_i^c \right\} \right]
    \nonumber \\ 
    &&- \frac{\sqrt{2}}{2}\left[\sum_c \left\{     (a b|c i) {}^{(2)}\kappa_ j^c + (a b|j c)           {}^{(2)}\kappa_ i^c\right\} - \sum_ k  \left\{ (j i|a k)
 {}^{(2)}\kappa_ k^b + (j i| k b)  {}^{(2)}\kappa_ k^a
    \right \}
    \right] \\
%
%
A^{(3)'''}_{a i, b j}  
    &=& 
    - \frac{1}{2}\Del{ab}\sum_{cdk} \binp{jc}{kd} {}^{(2)}\mptcoef{cd}{ik}
 - \frac {1}{2} \Del{ij}\sum_{ckl} \binp{kb}{lc} {}^{(2)}\mptcoef{ac}{kl} \\
%
A^{(3)''''}_{a i, b j}  &=& 
    \frac {1}{2} (\epsilon_b - \epsilon_j)
             \left ( \Del{ab} ~\rho^{(3)}_{ij}
                   - \Del{ij} ~\rho^{(3)}_{ba} \right ) \\
&=&                   
    - \frac{1}{4} (\epsilon_{b} - \epsilon_{j}) \left( \Del{ab} \sum_{kcd} \left( {}^{(1)}\kappa^{cd}_{ik}  {}^{(2)}\tilde{\kappa}^{cd}_{jk} + {}^{(2)}\kappa^{cd}_{ik}  {}^{(1)}\tilde{\kappa}^{cd}_{jk} \right) \right . \nonumber \\
    && \phantom{- \frac{1}{4} (\epsilon_{b}-\epsilon_{j})(} \left . 
    + \Del{ij} \sum_{klc} \left( {}^{(1)}\kappa^{ac}_{kl}  {}^{(2)}\tilde{\kappa}^{bc}_{kl} + {}^{(2)}\kappa^{ac}_{kl}  {}^{(1)}\tilde{\kappa}^{bc}_{kl} \right) \right)
\end{eqnarray}

\providecommand{\latin}[1]{#1}
\makeatletter
\providecommand{\doi}
  {\begingroup\let\do\@makeother\dospecials
  \catcode`\{=1 \catcode`\}=2 \doi@aux}
\providecommand{\doi@aux}[1]{\endgroup\texttt{#1}}
\makeatother
\providecommand*\mcitethebibliography{\thebibliography}
\csname @ifundefined\endcsname{endmcitethebibliography}  {\let\endmcitethebibliography\endthebibliography}{}

\end{document}